\documentclass[letterpaper, 10 pt, conference]{ieeeconf}  

\IEEEoverridecommandlockouts                              

\usepackage{amssymb}  

\title{\LARGE \bf
A Study on Social Robot Behavior in Group Conversation
}

\author{Tung Nguyen$^{1}$, Eric Nichols$^{1}$, and Randy Gomez $^{1}$
\thanks{$^{1}$Honda Research Institute Japan Co., Ltd., 8-1 Honcho, Wakoshi, Japan
        {\tt\small \{e.nichols,tung.nguyen,r.gomez\}@jp.honda-ri.com}}%
}

\begin{document}

\maketitle
\thispagestyle{empty}
\pagestyle{empty}

\begin{abstract}

Recently, research in human-robot interaction began to consider a robot’s influence at the group level. Despite the recent growth in research investigating the effects of robots within groups of people, our overall understanding of what happens when robots are placed within groups or teams of people is still limited. This paper investigates several key problems for social robots that manage conversations in a group setting, where the number of participants is more than two. In a group setting, the conversation dynamics are a lot more complicated than the conventional one-to-one conversation, thus, there are more challenges need to be solved. 

\end{abstract}

\section{INTRODUCTION}

In this paper, we discuss several key issues in terms of interaction management for a social robot that facilitates conversation in a group setting. In contrast to the conventional one-to-one conversation, which is the major research paradigm in human-robot interaction research, a group conversation, usually called multi-party dialog, involves three or more participants. As the number of interlocutors increases, the dynamics of the conversation also increase in terms of complexity, thus, making the facilitation task of the robot in a multi-party dialog very challenging. Despite the increase of recent research studying the effects of robots within groups of people, the overall understanding of what happens when robots participate in a conversation within groups of people and what the robot should do in this situation is still limited.

In principle, the handling of a multi-party dialog involves two phases: observation and interaction management. In the observation phase, the robot observes interactions among the other interlocutors to gather visual, acoustic, and verbal information of the current situation. Depending on the goal of the robot, the raw information from sensors are processed to extract the dynamics of the conversation such as who is the speaker, or who is being addressed, engagement level of each participants \cite{gupta2007resolving, gu2022says}. In the management phase, the robot performs facilitation behaviors accordingly to the perceived information and its pre-defined goal. Our work in this paper focuses on the phase of interaction management. The goal of a social robot that facilitates multi-party dialog is to maintain the flow of the conversation and make all the participants feel satisfied. 

Traum \cite{traum2004issues} discussed several key issues of a dialog system's interaction management in a group conversation setting, which includes: turn management, channel management, thread/conversation management, initiative management, and attention management. Each of these issues has been investigated in multiple existing studies. \cite{matsusaka2001modeling, bohus2011multiparty} proposed methods for handling turn taking in multi-party conversations. Matsuyama et al. \cite{matsuyama2015four} proposed a POMDP-based model that tackles the issue of initiative management to control the conversation flow among the participants. 

In this work, we discuss several important interaction management issues that a social robot needs to manage in a group conversation. The remainder of this paper is organized as follows. In Section \ref{sec:related_works}, we start by summarizing studies about facilitation in groups interaction. In Section \ref{sec:issues}, we discuss about key issues for robot to handle interaction management of group conversations effectively. Finally, we summarize and conclude this study in Section \ref{sec:conclusions}.

\section{RELATED WORKS}
\label{sec:related_works}
There are various research on specially situated facilitation agents in multiparty conversations. Ishizaki et al. \cite{ishizaki1998exploring} studied about differences in characteristics between dyadic (one-to-one) and multiparty conversations, making foundation for the later works. Interestingly, even though there are more research focus on robot behaviors in dyadic conversations, recent surveys show that humans are more likely to interact with a robot as a group \cite{gockley2006interactions, michalowski2006spatial}. Matsusaka et al. \cite{matsusaka2003conversation} were among the first to propose the utilization of a physical robot to participate in multiparty conversations. Traum \cite{traum2004issues} provided the first analysis of problems in handling conversation among group of interlocutors. Matsuyama et al. \cite{matsuyama2008designing} developed a multiparty quiz-game-type facilitation system for elderly. Another study reported the effectiveness of the existence of a robot in quiz-type group in terms of communication activation \cite{matsuyama2010psychological}. Sidner et al. \cite{sidner2004look} focused on engagement, a key factor to determine level of interaction among participants in conversation, which is defined as \textit{``the process by which two (or more) participants establish, maintain and end their perceived connection during interactions they jointly undertake''}. Based on this definition, they developed a robot that can maintain high level of engagement with human users using gaze behavior. 

Using the same definition as in Sidner's work, Bohus and Horvitz \cite{bohus2009models, bohus2010facilitating} evaluated the effectiveness of multimodality behavior model, which includes gaze, gesture, and speech, for a multiparty conversation facilitating agent. Kumar et al. \cite{kumar2011conversational} developed a model that selects appropriate dialogue action for facilitation in a tutoring scenario based on Bales’ Socio-Emotional Interaction Categories for text-based character agents. Recent studies show that robot can be effectively utilized to resolve conflicts among group members. For example, robots have demonstrated success in directly mediating children fighting over the same toy \cite{shen2018stop}. Martelaro et al. \cite{martelaro2015using} study showed that a robot intervening after a team member made a hostile remark toward the other helps increase the awareness of such conflict. However, the attempts of the robot to solve the conflict sometimes backfired. Thus, the behavior of the robot must be designed carefully in such cases to avoid heightening the level of the conflicts. 

\section{ISSUES IN GROUP CONVERSATION MANAGEMENT}
\label{sec:issues}
In this section, we describe the key issues of interaction management in multi-party conversation, which are: turn management, thread management, initiative management, and attention management. In order to handle multi-party dialog effectively and keep a high level of satisfaction among the human participants, a robot must be able to handle these key issues. The issues being discussed in this section also appear in dyadic conversation, however, they become a lot more complex when being considered in the context of multi-party conversations.

\subsection{Turn management}
Turn management refers to the problem of handling turn-taking in conversations, which means when to stop and when to start talking to the other participants. In a conventional setup, we force a rigid turn-taking manner, when a participant can only start talking when the other has finished their turn. However, there is another setup where the participant can ``barge-in'', which means to take the turn before the other finish talking. This is an important issue because turn is the basic unit of a conversation, thus, managing conversation turns is essential for controlling the flow of the conversation. 

In dyadic conversations, turn taking management is straight-forward because each participant takes a conversation turn after the other finish. As a result, the number of turns that each participant take during the conversation is usually balanced. In a multi-party dialog setting, when someone finish speaking, there are many participants that can take the turn to reply. As a result, there are cases where someone takes a lot of turns while the others have less. For a social robot that handles the conversation, such situation should be avoid, and the robot should intervene and prevents it from happening. Table \ref{tab:turn_management} shows an example of the robot taking turn and joins the conversation with the other participants. When the robot should take the turn is important in facilitating the conversation so that all participants do not feel being interrupted and satisfied. 

\begin{table}[h]
\caption{An example of a robot taking a turn in a multiparty conversation.}
\label{tab:turn_management}
\begin{center}
\begin{tabular}{|p{210pt}|}
\hline
Participants: A, B, and the robot\\
\hline
A: I am going to visit Italy this summer.\\
B: Oh, really? I think it's a beautiful country. Where in Italy are you going to stay?\\
A: I'm staying in Rome but I'll also visit Naples and Venice as well.\\
B: That sounds like an amazing trip!\\
A: I'm looking forward to try out all the Italian food there, too.\\
Robot: \textit{If you're staying in Rome, I recommend restaurant ABC. They have amazing carbonara there!}\\
\hline
\end{tabular}
\end{center}
\end{table}

\subsection{Thread management}
Thread management refers to the management of the conversation topics that are being discussed at the moment. In a one-to-one conversation, there are only two participants, thus, there can be only one topic exists at one time. Therefore, the management of topics is trivial in this case. However, in multi-party conversations, there can be multiple topics being discussed at the same time between multiple participants. For example, consider the case where we have five participants: A, B, C, D, and the robot. A and B are talking about football while C and D are talking about traveling. The robot needs to keep track of who is talking about which topic and gives the reply to the participants appropriately. An example of the robot handling topics of the conversation is shown in Table \ref{tab:thread_management}, when the robot joins the discussion of A and B, it talks about football and not about traveling.

\begin{table}[h]
\caption{An example of the robot handling topics in multi-party conversation.}
\label{tab:thread_management}
\begin{center}
\begin{tabular}{|p{210pt}|}
\hline
Participants: A, B, and the robot\\
Topic: football\\
\hline
A: Do you like football, B?\\
B: I like watching football but I don't play it. \\
A: Who is your favorite football player? I like Messi.\\
B: I don't have a favorite player but I like Real Madrid, they're my favorite football club since high school.\\
Robot: \textit{I like Real Madrid, too!}\\
\hline
\hline
Participants: C, D, and the robot\\
Topic: traveling\\
\hline
C: I'm going to Tokyo next month.\\
D: That's nice, what do you plan to do in Tokyo?\\
C: I like Japanese culture and want to visit the shrines and temples.\\
Robot: \textit{Let me see. Senso-ji in Asakusa is a really famous temple in Tokyo that you can visit.}\\
\hline
\end{tabular}
\end{center}
\end{table}

\subsection{Initiative management}
Initiative management concerns about which participant is currently setting the agenda for topics of discussions. In a dyadic conversation between robot and human, it is either robot-initiative (the robot starts the topics), human-initiative (the human user starts the topics), or mixed-initiative (both human and robot can start the topics). In multi-party dialogs, another type of initiative taking called cross-initiative can happen, where one participant asks another one to start a new topic for the conversation. 

For a robot that facilitates multi-party conversations, similar to turn management, it is important to make sure that there is no severe imbalance in terms of initiative taking. When necessary, the robot should take initiative and start a new topic of conversation. By doing this, we can keep all participants engaged and make the conversation goes smoothly.

\begin{table}[h]
\caption{An example of the robot handling initiative in multi-party conversation.}
\label{tab:thread_management}
\begin{center}
\begin{tabular}{|p{210pt}|}
\hline
Participants: A, B, and the robot\\
Topic: football\\
\hline
A: Do you like football, B?\\
B: I like watching football but I don't play it. \\
A: Who is your favorite football player? I like Messi.\\
B: I don't have a favorite player but I like Real Madrid, they're my favorite football club since high school.\\
...\\
\textit{conversation continues about football}\\
...\\
Robot: \textit{B, do you have any favorite kind of food?}\\
\hline
\end{tabular}
\end{center}
\end{table}

\subsection{Attention management}
Attention management in a dyadic conversation, the attention of the robot is always toward the only human user. However, in a multiparty dialog, there are multiple humans that the robot needs to give a response to while the robot can only give one response at a time. The task of who should the robot attend to at which scenario is important to facilitate the conversation successfully. The robot needs to assess the situations carefully and determine which participant to respond next. Failure to give appropriate attention to human participants can lead to dissatisfaction. In some cases, such as managing conversations among children, failure in attention management can lead to conflicts among the participants. Table \ref{tab:attention_management} shows an example where the robot needs to handle the attention from two participants.

\begin{table}[h]
\caption{An example of the robot handling attention in multi-party conversation.}
\label{tab:attention_management}
\begin{center}
\begin{tabular}{|p{210pt}|}
\hline
Participants: A, B, and the robot\\
\hline
A: Do you like football, B?\\
B: I like watching football but I don't play it. \\
A: Who is your favorite football player? I like Messi.\\
B: I don't have a favorite player but I like Real Madrid, they're my favorite football club since high school.\\
A: Robot, is there any interesting football match this weekend?\\
B: I am looking for a good Indian restaurant, can you give me some recommendation, robot?\\
Robot: There is Namaste restaurant near the station.\\
\hline
\end{tabular}
\end{center}
\end{table}

\section{CONCLUSIONS}
\label{sec:conclusions}
In this paper, we discuss several key issues that a social robot that facilitates multi-party conversations needs to take care of. In principle, to handle a multi-party dialogs, a robot needs to tackles problems in terms of turn management, thread management, initiative management, and attention management. Solving these challenges is essential for a robot to successfully facilitating a multi-party conversation.

In the future, we plan to analyse these key issues in more details and design behavior strategies for a social robot that manages multi-party conversations.

\addtolength{\textheight}{-12cm}   
\bibliographystyle{IEEEtran}
\bibliography{workshop_paper}

\end{document}